\documentclass[twocolumn]{aastex62}
\usepackage{amsmath}
\usepackage {autobreak}

\shorttitle{ALMA Polarization for AT2018cow}
\shortauthors{Huang, Shimoda, Urata, Toma et al.}
\begin{document}

\title{ALMA Polarimetry of AT2018cow}

\correspondingauthor{Kuyiun Huang, Yuji Urata}
\email{kuiyun@gmail.com, urata@g.ncu.edu.tw}

\author{Kuiyun Huang}
\affiliation{Center for General Education, Chung Yuan Christian University, Taoyuan 32023, Taiwan}

\author{Jiro Shimoda}
\affiliation{Frontier Research Institute for Interdisciplinary Sciences, Tohoku University, Sendai 980-8578, Japan}
\affiliation{Astronomical Institute, Tohoku University, Sendai, 980- 8578, Japan}

\author{Yuji Urata}
\affiliation{Institute of Astronomy, National Central University, Chung-Li 32054, Taiwan}

\author{Kenji Toma}
\affiliation{Frontier Research Institute for Interdisciplinary Sciences, Tohoku University, Sendai 980-8578, Japan}
\affiliation{Astronomical Institute, Tohoku University, Sendai, 980- 8578, Japan}

\author{Kazutaka Yamaoka}
\affiliation{Institute for Space-Earth Environmental Research (ISEE), Nagoya University, Furo-cho, Chikusa-ku, Nagoya, Aichi 464- 8601, Japan}
\affiliation{Division of Particle and Astrophysical Science, Graduate School of Science, Nagoya University, Furo-cho, Chikusa-ku, Nagoya, Aichi 464-8601, Japan}

\author{Keiichi Asada}
\affiliation{Academia Sinica Institute of Astronomy and Astrophysics, Taipei 106, Taiwan}

\author{Hiroshi Nagai}
\affiliation{National Astronomical Observatory of Japan, 2-21-1 Osawa, Mitaka Tokyo 181-8588, Japan}
\affiliation{Department of Astronomical Science, School of Physical Sciences, SOKENDAI (The Graduate University for Advanced Studies), Mitaka, Tokyo 181-8588, Japan}

\author{Satoko Takahashi}
\affiliation{Joint ALMA Observatory, Alonso de C{\'{o}}rdova 3107, Vitacura, Santiago, Chile}
\affiliation{NAOJ Chile Observatory, Alonso de C{\'{o}}rdova 3788, Oficina 61B, Vitacura, Santiago, Chile}
\affiliation{Department of Astronomical Science, School of Physical Sciences, SOKENDAI (The Graduate University for Advanced Studies), Mitaka, Tokyo 181-8588, Japan}

\author{Glen Petitpas}
\affiliation{Harvard-Smithsonian Center for Astrophysics, 60 Garden Street, Cambridge, Massachusetts 02138, USA}

\author{Makoto Tashiro}
\affiliation{Department of Physics, Saitama University, Shimo-Okubo, Saitama, 338-8570, Japan}


\begin{abstract}

We present the first radio polarimetric observations of a fast-rising
blue optical transient, AT2018cow. Two epochs of polarimetry with
additional coincident photometry were performed with the Atacama Large
Millimeter/submillimeter Array (ALMA). The overall photometric results
based on simultaneous observations in the 100 and 230 GHz bands
are consistent with the non-thermal radiation model reported by
\citet{ho19} and indicate that the spectral peaks ($\sim110$ GHz at
the first epoch and $\sim67$ GHz at the second epoch) represent
the synchrotron self-absorption frequency.  The non-detection of linear
polarization with $<$0.15\% in the 230 GHz band at the phase when the
effect of synchrotron self-absorption was quite small in the band may
be explained by internal Faraday depolarization with high
circumburst density and strong magnetic field.  This result supports
the stellar explosion scenario rather than the tidal disruption model.
The maximum energy of accelerating particles at the shocks of
AT2018cow-like objects is also discussed.

\end{abstract}


\section{Introduction} \label{sec:intro}

A luminous transient, AT2018cow, was discovered near the galaxy CGCG
137-068 ($z$ =0.0141) at 2018-06-16 10:35:02 UT \citep{smartt18}.
High luminosity in various wavelengths, featureless
hot black-body spectra, and long-lived radio emission revealed that
AT2018cow is an unusual transient
\citep{rivera18,prentice18,kuin19,ho19}.
Panchromatic approaches suggested the presence of the central engine
of high-energy emission radiated through equatorial-polar asymmetric
low-mass ejecta in a dense medium, and the progenitor of a
low-mass H-rich star or blue supergiant star \citep{margutti19, soker19, ho19}.
A scenario was also proposed in which a star disrupted by an
intermediate black hole produced AT2018cow
\citep{perley19,kuin19}. However, the large environment density
concluded by \citet{margutti19, ho19} made the tidal disruption scenario
unlikely and indicated a stellar explosion hypothesis.  The host
galaxy observation with HI 21cm mapping demonstrated that AT2018cow
lies within an asymmetric ring of high column density, which indicates
the formation of massive stars, supporting the stellar explosion
scenario of AT2018cow \citep{roychowdhury19}.  \citet{lyutikov19}
built an electron-capture collapse model following a merger of white
dwarfs one of which is a massive ONeMg white dwarf.

Polarimetry may be another key to investigating the circumstances of stellar
explosion objects, such as density, magnetic field, and
turbulence. Moreover, the study of particle acceleration at shocks
associated with the objects could be equally interesting. 
%
%
For SN~1987A \citep{zanardo18} and Kepler's supernova remnant \cite[SNR,][]{delaney02} as
examples, spatially-resolved linear polarizations of radio synchrotron
emissions were observed with local polarization degrees of $\sim10$\%.  The
local polarization angles of both objects imply a radially oriented
magnetic field. The polarization degree for integrated Stokes
parameters over all emission regions is a few per cent.  Such radial
orientations and sizable polarization degrees are ubiquitously
observed in young SNRs (such as the freely expanding phase to early Sedov
phase, e.g. \citealt{milne87}; \citealt{dickel91} for Tycho's SNR;
\citealt{reynoso13} for SN~1006) and could be explained by
magnetohydrodynamic turbulence resulting from the interaction between
the shock wave and density fluctuations pre-existing in the upstream
medium (i.e. stellar wind and/or interstellar medium,
\citealt{inoue13}).
As for the early stages of radio supernovae, however, the density and
magnetic field strength in the shocked region can be so high that
the Faraday rotation effect is strong.  Then the emissions from
different parts in the shocked region have different polarization
angles, which lead to suppression of the net linear polarization,
i.e., the internal Faraday depolarization.  The non-detection of
linear polarization at 1.7$-$8.4 GHz in SN 1993J is explained by this effect
\citep{bietenholz03}.

In this paper, we report the radio polarimetry of AT2018cow using the
Atacama Large Millimeter/submillimeter Array (ALMA) in the 100 GHz and
230 GHz bands. In this millimeter wavelength range, the Faraday effect
is weaker than the centimeter radio band. Based on two epochs of ALMA
observations, the scenarios of a progenitor accompanied by a
dense circumstellar medium are examined.
MJD 58285 (2018-06-16 00:00:00 UT) is used as $T_{0}$, which is between
the last non-detection (MJD 58284.13) and the date of discovery (MJD
58285.441). The date is the same $T_{0}$ used in \citep{perley19, ho19}.

\section{Observations} \label{sec:obs}

Two epochs of ALMA observations were executed as part of Director's
Discretionary Time (DDT) during Cycle 5 (2017.A.00046.T; PI K. Huang)
using both the 12-m antenna array and Atacama Compact Array (ACA). The
first epoch of radio linear polarimetry was performed at 97.5 GHz
(i.e. Band3) starting at 27 June 2018 01:04 UT (midpoint $T_{0}$=11.1
d, here after epoch1).
Coincident 230-GHz band (i.e. Band6) observations were also performed
with the ACA. Because our quick-look photometry using the ACA data
exhibited the brightness sufficient for polarimetry and positive
power-law index by fitting with $f_{\nu}\propto\nu^{\beta}$, we
decided to switch the frequency from Band3 to Band6 to perform
polarimetry above the spectral peak. Hence, the second epoch of
polarimetry was executed at the 230 GHz band using the 12-m antenna
array on 3 July 2018 UT (midpoint $T_{0}$=17.1 d, here after
epoch2). The coincident photometry at 97.5 GHz was also performed
using the ACA.
For the 12-m antenna array, the bandpass and flux were calibrated
using observations of J1550+0527, and J1606+1814 was used for the
phase calibration. Polarization calibration was performed by
observations of J1642+3948. Regarding ACA observations, J1337-1257 and
J1517-2422 were utilized for the bandpass and flux calibrations. The phase
calibrations were performed using observations of J1540+1447,
J1613+3412, and J1619+2247.

\begin{figure}
\epsscale{1.15}
\plotone{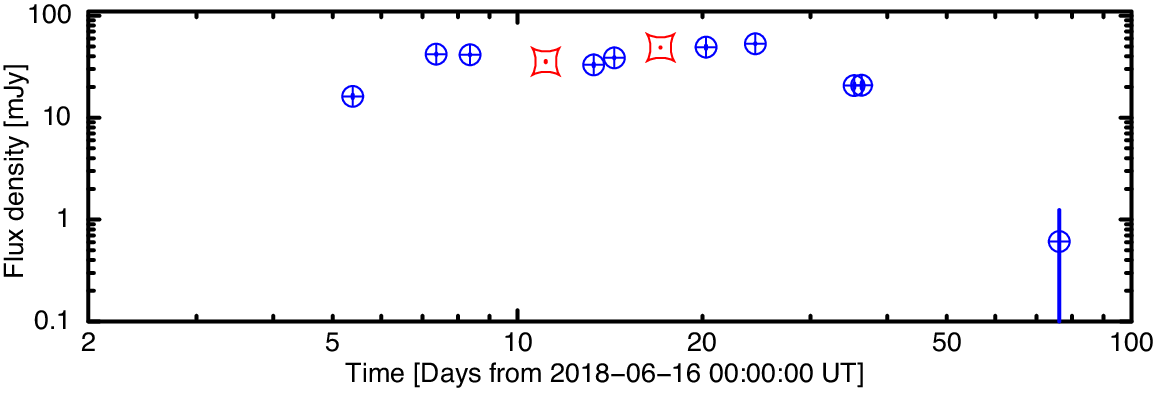}
\caption{AT2018cow light curve in the submillimeter band ($\sim$230 GHz). The red box points indicate the photometric results of the ALMA and the blue circle points show the monitoring results reported by \citet{ho19}.}
\label{lc}
\end{figure}

\section{Analysis and Results} \label{sec:result}
The raw data of ALMA were reduced at the East Asian ALMA Regional
Center (EA-ARC) using CASA (version 5.1.1) \citep{casa}.
We further performed interactive CLEAN deconvolution imaging
\citep{clean1,clean2} with self-calibration for the data obtained by
the 12-m antenna array. The Stokes $I$, $Q$, and $U$ maps were CLEANed with
an appropriate number of CLEAN iterations after the final round of
self-calibration.
The results of photometry and polarimetry are summarized in Table
\ref{almapol}.
Regarding polarimetry, the 3-$\sigma$ upper limits were derived based on
the non-detections in Q and U maps.
Because the depolarization between the source and observation site is
negligible for the point source (i.e., transients) in this millimeter
band \citep{brentjens05}, the values of $<0.10\%$ in the 97.5-GHz band
and $<0.15\%$ in the 233-GHz band describe the intrinsic origin.


To describe the phase of the polarization observation, the 
photometric measurements in the entire Band6 frequency range were plotted, together
with the 230-GHz monitoring data \citep{ho19}. As shown in Figure
\ref{lc}, the $\sim$230-GHz light curves indicate that our polarimetric
measurements were performed around the brightest plateau phase with
significant variabilities.

\begin{figure*}
\epsscale{.9}
\plotone{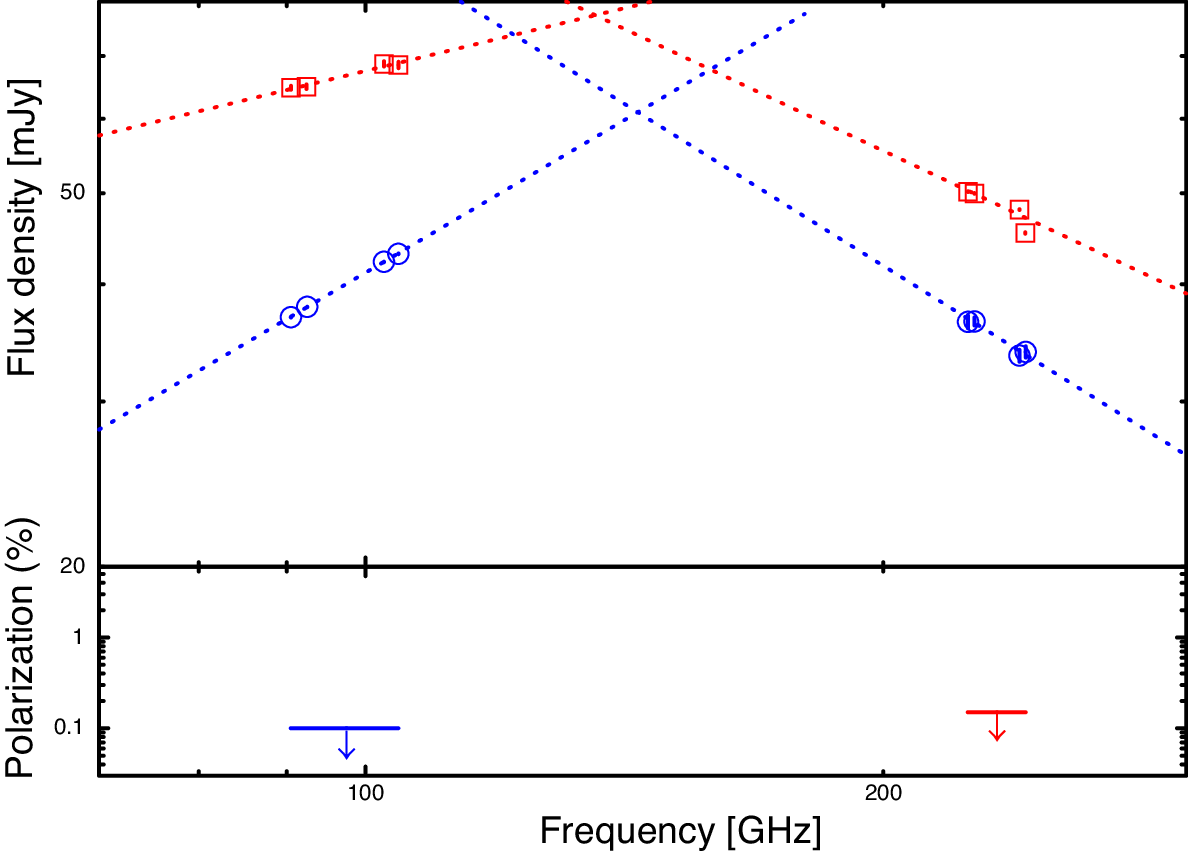}
\caption{SED and polarization using ALMA Band3 and Band6 data taken at 11.1 (blue circle points and arrow) and 17.1 (red box points and arrow) d. The blue and red dotted lines indicate the best fitted simple power-law functions.}
\label{almased}
\end{figure*}

The photometric measurements in each of the spectral windows
(SPWs) of Band3 and Band6 were fitted with a simple power-law function
(i.e. $f_{\nu}\propto\nu^{\beta}$).  These fittings yield
$\beta_{E1B3}=1.080\pm0.007$ ($\chi^2/ndf$=1.39 with number of degree of freedom, $ndf$=2) for
epoch1 with Band3, $\beta_{E1B6}=-1.15\pm0.16$ ($\chi^2/ndf$=0.62 with ndf=2) for epoch1 with Band6, $\beta_{E2B3}=0.44\pm0.05$
($\chi^2/ndf$=1.00 with $ndf$=2) for epoch2 with Band3, and
$\beta_{E2B6}=-0.86\pm0.29$ ($\chi^2/ndf$= 280 with $ndf$=2) for
epoch2 with Band6.
Large scatter were observed at 242 GHz (i.e. the highest frequency in Band6) for epoch2, which
may be related to the significant variabilities as shown in Figure \ref{lc}.  
The same fitting was therefore performed by excluding the data, and
$\beta_{E2B6}=-0.65\pm0.02$ ($\chi^2/ndf$=1.25 with $ndf$=1) was obtained.
As shown in Figure \ref{almased}, the fitting basically describes the
spectral energy distribution (SED) and indicates that the spectral
peak frequency, $\nu_{p}$ is located at $\sim$140 GHz.  Hence, the
polarimetric measurements on epoch1 and epoch2 were performed below
and above the spectral peak, respectively.
%

\section{Discussion} \label{sec:discuss}

\subsection{Spectral Flux Distribution}

\begin{figure*}
\epsscale{.9}
\plotone{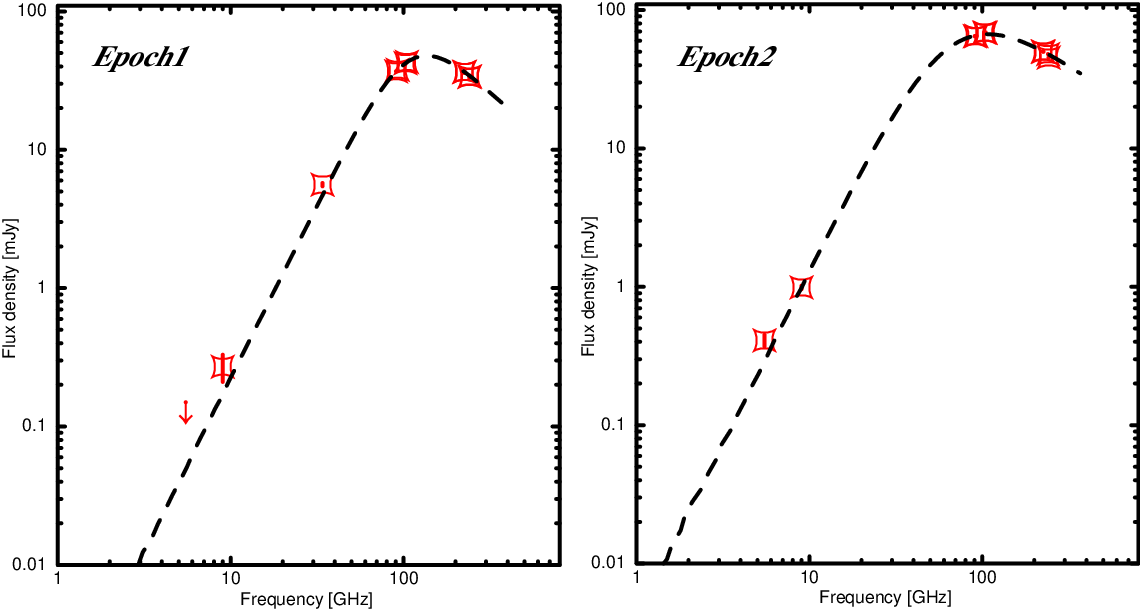}
\caption{SED of the AT2018cow at 11.1 d (i.e., epoch1, left) and 17.1 d (i.e., epoch2, right) using ALMA with the ATCA data taken by \citet{ho19}. Time differences of ATCA observations were $\Delta t=-$0.6 d for epoch1 and
$\Delta$t=0.4 d for epoch2, respectively. The dashed lines indicate the best fit smoothly connected broken power-law functions with the spectral peak frequencies of $\sim$110 GHz at epoch1 and $\sim$67 GHz at epoch2.}
\label{radiosed}
\end{figure*}

The observed radio light curves and time-resolved spectra of AT2018cow
may be interpreted as the synchrotron emission of relativistic
non-thermal electrons produced at an adiabatic strong shock that
freely expands in an ionized medium at a non-relativistic speed
\citep{ho19,margutti19}. This emission model is widely applicable for
radio supernovae \citep{chevalier98}.
Considering the smooth connection of two power-law spectra,
the temporal evolution of the spectral indices in Band 3 may be
consistent with the spectral modeling presented by \citet{ho19}.
The smooth broken power-law fitting was performed, including ATCA data
taken at the similar epochs ($\Delta t=-$0.6 d for epoch1 and
$\Delta$t=0.4 d for epoch2 \citep{ho19}). 
The smooth fitting with wider spectral frequency coverage is also
reasonable to characterize the spectral peak frequency as the method
is applied for various analyses such as GRB prompt emissions \citep[e.g.][]{band}.
Because the significant variabilities were observed (Figure \ref{lc}),
we excluded the data taken by the Submillimeter Array (SMA). For this fitting, the spectral index of the lower-frequency
side was fixed as $\beta_{low}=2.5$ (reported by
\citet{ho19}), and the higher-frequency side was fixed as
$\beta_{high}=-1.15$ for epoch1 and $\beta_{high}=-0.86$ for epoch2.
The fitting yields the spectral peak frequency, $\nu_{p}$ =109.8$\pm$0.5 GHz ($\chi^2/ndf$=7.3 with $ndf=$7) at epoch1 and $\nu_{p}$ = 67.4$\pm$1.6 GHz ($\chi^2/ndf$=7.6 with $ndf=6$) at epoch2\footnote{The differences between our deduced spectral peak frequencies and $\sim$100 GHz at 22 day estimated by \citet{ho19} may be caused by their analysis for narrow frequency range and power-law index measurement ($-1.06\pm0.01$) using interpolated SMA data (between 20 d and 24 d). Much flatter spectral index may be reasonable to explain the spectral excess of their measurement with 671 GHz at 23 d.}.
The larger
reduced $\chi^2$ may be caused by the epoch differences. 
As show in Figure \ref{radiosed}, the best-fit functions basically
describe the SED.  The temporal evolution of the spectral peak
frequency is also characterized as $\nu_{p} \propto t^{-1.1}$, which
is consistent with that of the theoretical model for the synchrotron
self-absorption frequency \citep{chevalier98}.
Therefore, we concluded that the spectral peak frequency represents the
synchrotron self-absorption frequency, and the effect of self-absorption is quite small for the polarization measurement
with in the 233-GHz band at epoch2.

For further discussion in \S4.2, the theoretical
analysis of \citet{ho19} is followed for the estimated values of the radius of the
shock, magnetic field strength in the shocked region, shock speed, and
number density of the shocked region at $T \simeq 22\;$d as

\begin{eqnarray}
  &&R \simeq 7 \times 10^{15}\;\left(\frac{\epsilon_e}{\epsilon_B}\frac{f}{0.5}\right)^{\frac{-1}{19}} \left(\frac{F_p}{94\;{\rm mJy}}\right)^{\frac{9}{19}} \nonumber\\
  &&~~~~~~~~~~~\times\left(\frac{D}{60\;{\rm Mpc}}\right)^{\frac{18}{19}} \left(\frac{\nu_p}{100\;{\rm GHz}}\right)^{-1}\;{\rm cm},\\
  &&B \simeq 6\;\left(\frac{\epsilon_e}{\epsilon_B}\frac{f}{0.5}\right)^{\frac{-4}{19}} \left(\frac{F_p}{94\;{\rm mJy}}\right)^{\frac{-2}{19}} \nonumber \\
  &&~~~~~~~~~~~\times\left(\frac{D}{60\;{\rm Mpc}}\right)^{\frac{-4}{19}} \left(\frac{\nu_p}{100\;{\rm GHz}}\right)\;{\rm G}, \\
  &&\frac{v}{c} \simeq 0.1\;\left(\frac{\epsilon_e}{\epsilon_B}\frac{f}{0.5}\right)^{\frac{-1}{19}} \left(\frac{F_p}{94\;{\rm mJy}}\right)^{\frac{9}{19}} \nonumber\\
  &&~~~~~~~~\times\left(\frac{D}{60\;{\rm Mpc}}\right)^{\frac{18}{19}} \left(\frac{\nu_p}{100\;{\rm GHz}}\right)^{-1} \left(\frac{T}{22\;{\rm day}}\right)^{-1}, \\
  &&n_e \simeq 1 \times 10^5\;\frac{1}{\epsilon_B}\left(\frac{\epsilon_e}{\epsilon_B}\frac{f}{0.5}\right)^{\frac{-6}{19}} \left(\frac{F_p}{94\;{\rm mJy}}\right)^{\frac{-22}{19}} \nonumber \\
  &&~~~\times\left(\frac{D}{60\;{\rm Mpc}}\right)^{\frac{-44}{19}} \left(\frac{\nu_p}{100\;{\rm GHz}}\right)^{4} \left(\frac{T}{22\;{\rm day}}\right)^2 \;{\rm cm}^{-3},
\end{eqnarray}
where $\epsilon_e$ and $\epsilon_B$ are the fractions of thermal
energy at the shocked region that are carried by the non-thermal
electrons and the magnetic field, respectively, and $f$ is the filling
factor of the emission region in the sphere with radius $R$.  

\subsection{Polarization}
As introduced in \S1, the polarization degree of AT2018cow without the
Faraday effect is expected to be a few percent, which is similar to
other stellar explosions\footnote{The optical linear polarizations were detected \citep{optpol}
when the thermal radiation dominated in the optical range
\citep{perley19,margutti19}.  Hence, the values are not appropriate to
refer to as the polarization degree without the Faraday depolarization
effect.}.
The non-detection of linear polarization (especially $<$0.15\% in the
233-GHz band at epoch2) in AT2018cow may be explained by internal
Faraday depolarization, because $n_e$ and $B$ are so high. 
The result supports the stellar explosion scenario rather than the
tidal disruption scenario\footnote{In the tidal disruption scenario, the external shock which propagates in the
interstellar medium will have $n_{e}\sim 1\; {\rm cm}^{-3}$, $B \sim 1\; \mu$G \citep{perley19,gaensler05}, $R < c \times 22\;$ day. These lead to $\tau_{V} < 3\times10^{-7} \ll 1$ in Eq. (5) even for $N=1$, and then a sizable linear polarization may be detected.}.

In this scenario, we can derive a lower limit of the coherence length
of the turbulent magnetic field in the shocked region.  Supposing that the
turbulent magnetic energy peaks at the maximum coherence length scale
$\ell_{\rm M}$, which is observationally implied in Tycho's SNR
\citep{shimoda18}, we obtain the Faraday depth as

\begin{align}
\begin{autobreak}
\tau_V = \frac{e^3}{\pi m_e^2 c^2} n_e \frac{B}{\sqrt{N}} R \nu^{-2},
\end{autobreak}
\end{align}
where $N \sim R/\ell_{\rm M}$.
The condition $\tau_V > 1$ at $\nu \sim 100\;$GHz gives
  \begin{eqnarray} 
  \ell_{\rm M} > &&\ell_V \equiv \left(\frac{\pi m_e^2 c^2}{e^3}\frac{\nu^{2}}{n_{e}B\sqrt{R}}\right)^{2} \nonumber \\
  &&= 2 \times 10^6\;\left(\frac{n_e}{3 \times 10^5\;{\rm cm}^{-3}}\right)^{-2} \left(\frac{B}{6\;{\rm G}}\right)^{-2} \nonumber \\
  &&~~~~~~~~~~\times\left(\frac{R}{7\times10^{15}\;{\rm cm}}\right)^{-1} \left(\frac{\nu}{100\;{\rm GHz}}\right)^4 \;{\rm cm}.
 \end{eqnarray}
The observation of Tycho's SNR indicates $\ell_M \sim R/10$ \citep{shimoda18}. If this relation is valid for AT2018cow, the values of $n_e$, $B$, and $R$ estimated by \citet{ho19} and \citet{margutti19}(Eqs. 1, 2, and 4) satisfy Eq. (6).

The lower limit on $\ell_M$ leads to the lower limit on the maximum energy of accelerating particles at the shock.
In the first-order Fermi acceleration, which is assumed by \citet{ho19}, energetic particles are scattered through interactions
with the turbulent magnetic-field to go back and forth between
upstream and downstream of the shock, and then gain energies at every
reciprocation \citep{bell78,blandford78}.  The particles
experience large angle scattering if they resonantly interact with
magnetic disturbances with a scale length comparable to their gyro
radius, i.e. a pitch-angle scattering \citep{jokipii66}.  When the gyro
radius of accelerated particles becomes larger than the maximum
coherence length scale $\ell_{\rm M}$, the particle is no longer
efficiently scattering and escapes from the shock.  Thus, we obtain
the maximum energy of accelerating particles as
\begin{eqnarray}
  E_{\rm max} &&> eB \ell_V \nonumber \\
  &&\simeq 3\; \left(\frac{n_e}{3 \times 10^5\;{\rm cm}^{-2}}\right)^{-2} \left(\frac{B}{6\;{\rm G}}\right)^{-1} \nonumber \\
  &&~~~~\times\left(\frac{R}{7\times10^{15}\;{\rm cm}}\right)^{-1} \left(\frac{\nu}{100\;{\rm GHz}}\right)^4 \;{\rm GeV}.
\end{eqnarray}
This argument is consistent with the model in which the relativistic
non-thermal electrons are produced by the shock in AT2018cow.

The strong $\nu$ dependence of the lower limit on $E_{\rm max}$ should
be emphasized. If one can perform polarimetric observation of such
kinds of stellar explosions at higher frequencies, a stricter limit on $E_{\rm max}$ can
be obtained.
The origin of the PeV energy cosmic-rays is unknown. By polarimetry at
a higher $\nu$ (i.e. $\sim$THz), we could examine whether
AT2018cow-like objects are the origin of PeV cosmic-rays.

\acknowledgments

This paper makes use of the following ALMA data:
ADS/JAO.ALMA\#2017.A.00046.T. ALMA is a partnership of ESO
(representing its member states), NSF (USA) and NINS (Japan), together
with NRC (Canada), MOST and ASIAA (Taiwan), and KASI (Republic of
Korea), in cooperation with the Republic of Chile. The Joint ALMA
Observatory is operated by ESO, AUI/NRAO and NAOJ.
This work is supported by the Ministry of Science and Technology of
Taiwan grants MOST 105-2112-M-008-013-MY3 (Y.U.) and
106-2119-M-001-027 (K.A.). This work is also supported by JSPS
Grants-in-Aid for Scientific Research No. 18H01245 (K.T.). We thank
EA-ARC, especially Pei-Ying Hsieh for support in the ALMA
observations.  Y.U, K. Y. H, and K. A. also thank Ministry of
Education Republic of China.

%

\vspace{5mm}
\facilities{ALMA}


\software{CASA \citep{casa}  
          }

\begin{table}[hbtp]
\caption{ALMA Observing Log}
\label{almapol}
\centering
\begin{tabular}{cccccccc}
\hline
\multicolumn{8}{l}{Epoch1: 2018-06-27 01:04-04:43, $T_{0}$=11.1 d (midpoint)} \\ 
Instruments & SPW & Band [GHz] & Pol. [\%] & P.A [deg] & I flux [mJy] & Q flux [mJy] & U flux [mJy] \\
\hline
12m & 0,1,2,3 &  97.5 & $<0.10$ & $-$ & 39.629$\pm$0.046 & 0.010 (rms) & 0.010 (rms)\\
    & 0       &  90.5 & $<0.18$ & $-$ & 36.886$\pm$0.029 & 0.016 (rms) & 0.014 (rms)\\
    & 1       &  92.5 & $<0.17$ & $-$ & 37.823$\pm$0.029 & 0.016 (rms) & 0.015 (rms)\\
    & 2       & 102.5 & $<0.15$ & $-$ & 42.243$\pm$0.030 & 0.015 (rms) & 0.015 (rms)\\
    & 3       & 104.5 & $<0.17$ & $-$ & 43.087$\pm$0.036 & 0.017 (rms) & 0.017 (rms)\\
\hline
\multicolumn{8}{l}{Epoch1: 2018-06-27 01:00-04:05, $T_{0}$=11.1 d (midpoint)}\\ 
Instruments & SPW & Band [GHz] & Pol. [\%] & P.A [deg] & I flux [mJy] & Q flux [mJy] & U flux [mJy] \\
\hline
ACA & 4,6,16,18 & 233.0 & $-$ & $-$ & 35.49$\pm$0.48 & $-$ & $-$ \\
    & 4         & 224.0 & $-$ & $-$ & 36.47$\pm$0.57 & $-$ & $-$ \\
    & 6         & 226.0 & $-$ & $-$ & 36.51$\pm$0.37 & $-$ & $-$ \\
    & 16        & 240.0 & $-$ & $-$ & 33.57$\pm$0.49 & $-$ & $-$ \\
    & 18        & 242.0 & $-$ & $-$ & 33.88$\pm$0.47 & $-$ & $-$ \\
\hline
\multicolumn{8}{l}{Epoch2: 2018-07-03 00:39-02:06, $T_{0}$=17.1 d (midpoint)}\\ 
Instruments & SPW & Band [GHz] & Pol. [\%] & P.A [deg] & I flux [mJy] & Q flux [mJy] & U flux [mJy] \\
\hline
ACA & 4,6,8,10 &  97.5 & $-$ & $-$ & 68.18$\pm$0.44 & $-$ & $-$ \\
    & 4        &  90.5 & $-$ & $-$ & 64.74$\pm$0.31 & $-$ & $-$ \\
    & 6        &  92.4 & $-$ & $-$ & 64.93$\pm$0.37 & $-$ & $-$ \\
    & 8        & 102.5 & $-$ & $-$ & 68.66$\pm$0.45 & $-$ & $-$ \\
    & 10       & 104.5 & $-$ & $-$ & 68.47$\pm$0.53 & $-$ & $-$ \\
\hline
\multicolumn{8}{l}{Epoch2: 2018-07-03 00:43-04:05, $T_{0}$=17.1 d (midpoint)}\\ 
Instruments & SPW & Band [GHz] & Pol. [\%] & P.A [deg] & I flux [mJy] & Q flux [mJy] & U flux [mJy] \\
\hline
12m & 0,1,2,3 & 232.9 & $<0.15$ & $-$ & 48.755$\pm$0.047 & 0.016 (rms) & 0.018 (rms)\\ 
    & 0       & 224.0 & $<0.23$ & $-$ & 50.186$\pm$0.052 & 0.025 (rms) & 0.028 (rms)\\
    & 1       & 226.0 & $<0.22$ & $-$ & 49.975$\pm$0.046 & 0.026 (rms) & 0.025 (rms)\\
    & 2       & 240.0 & $<0.25$ & $-$ & 48.023$\pm$0.050 & 0.027 (rms) & 0.029 (rms)\\
    & 3       & 242.0 & $<0.31$ & $-$ & 45.348$\pm$0.086 & 0.033 (rms) & 0.033 (rms)\\
\hline
\hline
\end{tabular}
\end{table}

\end{document}